\documentclass[fleqn,usenatbib]{mnras}
\usepackage{newtxtext,newtxmath}
\usepackage[T1]{fontenc}

\DeclareRobustCommand{\VAN}[3]{#2}
\let\VANthebibliography\thebibliography
\def\thebibliography{\DeclareRobustCommand{\VAN}[3]{##3}\VANthebibliography}

\usepackage{graphicx}	
\usepackage{amsmath}	
\usepackage{upgreek}    
\usepackage{threeparttable}
\usepackage{tabulary}

\newcommand{\DMunits}{\,pc\,cm$^{-3}$}
\newcommand{\RMunits}{\,rad\,m$^{-2}$}
\newcommand{\Swin}{Centre for Astrophysics and Supercomputing, Swinburne University of Technology, P.O. Box 218, Hawthorn, VIC 3122, Australia}
\newcommand{\GWDC}{Gravitational Wave Data Centre, Swinburne University of Technology, P.O. Box 218, Hawthorn, VIC 3122, Australia}
\newcommand{\sydney}{Sydney Institute for Astronomy, School of Physics, University of Sydney, Camperdown, NSW 2006, Australia}

\newcommand{\ATNF}{Australia Telescope National Facility, CSIRO Astronomy and Space Science, P.O. Box 76, Epping, NSW 1710, Australia}
\newcommand{\NRL}{Space Science Division, Naval Research Laboratory, Washington, DC 20375-5352, USA}
\newcommand{\berkeley}{Department of Astronomy, University of California Berkeley, Berkeley, CA 94720, USA}
\newcommand{\NRCC}{National Research Council of Canada, Herzberg Institute of Astrophysics, Dominion Radio Astrophysical Observatory, P.O. Box 248, Penticton, BC V2A 6J9, Canada}
\newcommand{\ozgrav}{ARC Centre of Excellence for Gravitational Wave Discovery (OzGrav), Australia}
\newcommand{\NRAO}{National Radio Astronomy Observatory, 1003 Lopezville Rd, Socorro, NM 87801, USA}
\newcommand{\SETI}{SETI Institute, Mountain View, CA 94043, USA}

\title[Repetition from FRB\,190711]{Extremely band-limited repetition from a fast radio burst source}

\author[Kumar et al.]{P.~Kumar$^{1}$\thanks{E-mail: pravirkumar@swin.edu.au},
R.~M.~Shannon$^{1,2}$,
C.~Flynn$^{1}$,
S.~ Os{\l}owski$^{1,3}$,
S. Bhandari$^{4}$,
C.~K.~Day$^{1,4}$,
A.~T.~Deller$^{1}$,
W.~Farah$^{1,5}$,
\newauthor
J.~F.~Kaczmarek$^{6}$,
M.~Kerr$^{7}$,
C.~Phillips$^{4}$,
D.~C.~Price$^{1,8}$,
H.~Qiu$^{9,4}$ and
N.~Thyagarajan$^{10}$\thanks{N.~Thyagarajan is a Jansky Fellow of the National Radio Astronomy Observatory.}
\\
$^{1}$\Swin\\
$^{2}$\ozgrav\\
$^{3}$\GWDC\\
$^{4}$\ATNF\\
$^{5}$\SETI\\
$^{6}$\NRCC\\
$^{7}$\NRL\\
$^{8}$\berkeley\\
$^{9}$\sydney\\
$^{10}$\NRAO
}

\date{Accepted 2020 October 30. Received 2020 October 14; in original form 2020 September 4}

\pubyear{2020}

\begin{document}
\label{firstpage}
\pagerange{\pageref{firstpage}--\pageref{lastpage}}
\maketitle

\begin{abstract} 
The fast radio burst (FRB) population is observationally divided into sources that have been observed to repeat and those that have not. There is tentative evidence that the bursts from repeating sources have different properties than the non-repeating ones. In order to determine the occurrence rate of repeating sources and characterize the nature of repeat emission, we have been conducting sensitive searches for repetitions from bursts detected with the Australian Square Kilometre Array Pathfinder (ASKAP) with the 64-m Parkes radio telescope, using the recently commissioned Ultra-wideband Low (UWL) receiver system, over a band spanning 0.7--4.0~GHz. We report the detection of a repeat burst from the source of FRB\,20190711A. The detected burst is 1~ms wide and has a bandwidth of just 65 MHz. We find no evidence of any emission in the remaining part of the 3.3~GHz UWL band. While the emission bandwidths of the ASKAP and UWL bursts show $\nu^{-4}$ scaling consistent with a propagation effect, the spectral occupancy is inconsistent with diffractive scintillation. This detection rules out models predicting broad-band emission from the FRB\,20190711A source and puts stringent constraints on the emission mechanism. The low spectral occupancy highlights the importance of sub-banded search methods in detecting FRBs.
\end{abstract}




\section{Introduction}\label{sec:intro}
Fast radio bursts (FRBs) are providing new ways to study high energy processes and probe the distribution of matter in the Universe. The cosmological origin of FRBs has been confirmed \citep{Chatterjee:2017, Bannister:2019_localization, Prochaska:2019, Ravi:2019_localization, Bhandari:2020,Bhandari:2020_191001, Marcote:2020} and localized bursts have been successfully utilized to measure the baryon density of the low-redshift Universe \citep{Macquart:2020}. Nevertheless, the physical mechanism behind these perplexing, bright, millisecond-duration radio transient events still remains unknown \citep{Cordes:2019, Petroff:2019}.

Over the last five years, the number of detected FRBs has rapidly increased, with now about 130 published FRB sources on the Transient Name Server (TNS\footnote{\url{https://wis-tns.weizmann.ac.il/}; visited 2020 August 22.}). There are at present 20 sources known to repeat \citep{Spitler:2016, CHIME_repeaters19_1, CHIME_repeaters19_2, Kumar:2019, Fonseca:2020}. One of them, the source of FRB\,20180916B, shows periodic activity \citep{CHIME20}, and there is tentative evidence for periodic activity from another \cite[FRB\,20121102A;][]{Rajwade:2020}. More recently, the discovery of FRB-like pulses of radio emission associated with the magnetar SGR\,1935+2154 in the Milky Way \citep{Bochenek:2020, CHIME:2020:Magnetar} increased the credibility that some FRBs can be produced by magnetar-like progenitors at extragalactic distances.

The existence of repeating FRBs clearly indicates that a significant fraction of FRBs are not caused by cataclysmic events as has been speculated for one-off (apparently non-repeating) FRB sources. Initially thought to be coming from multiple progenitor populations of FRBs \citep{Palaniswamy:2018} due to the diversity in observed properties and thousands of hours of follow-up time spent with no repetitions, the dichotomy \citep{Caleb:2018} is now blurring in favour of sources having different observational repetition rates \citep{Caleb:2019, Connor:2020, James:2020}, and instrumental sensitivity biases affecting the ability to confirm repetition \citep{Kumar:2019, Lu:2020}. An analysis of volumetric burst rate also suggests that most, if not all, FRBs are produced by repeating progenitors \citep{Ravi:2019}. However, whether all FRB sources repeat remains an open question.

FRBs have been detected over a broad range of frequencies; the highest frequency detection is from the FRB\,20121102A source at 8~GHz \citep{Gajjar:2018}, and the lowest is from the periodically active source FRB\,20180916B \citep{Chawla:2020, Pilia:2020} and the FRB\,20200125A \citep{Parent:2020} at 300~MHz. Even so, any study of the spectral occupancy is not well constrained because of the limited band extent of the observing telescopes. The spectral index and spectral shape of FRB emission can be used to constrain and test proposed models of burst progenitors \citep{Platts:2019}. One strategy is to target repeating sources with multiple telescopes simultaneously. Many such observations have been conducted for FRBs\,20121102A and 20180916B \citep{Gourdji:2019, Hessels:2019, Houben:2019}, mostly resulting in non-detections of any coincident emission at two different frequencies, with two exceptions \citep{Law:2017, Chawla:2020}. In the case of \citet{Chawla:2020}, the coincident detections were observed in adjacent frequency bands of the Robert C. Byrd Green Bank Telescope (GBT) and the Canadian Hydrogen Intensity Mapping Experiment (CHIME) radio telescope. All these efforts have led to the conclusion that the rate of burst detection strongly depends on the radio frequency band being observed \citep{Majid:2020}. It is still not clear if these band-limited emissions are intrinsic to the source or caused by propagation effects \citep{Cordes:2017}.

While simultaneous observations are crucial to understanding the spectral nature of FRBs, there are several challenges in undertaking multifrequency campaigns and interpreting their results. Significant coordination and strategy are needed for the observing proposal to conduct such programs \citep{Law:2017}. Also, differences in sensitivity, radio frequency interference (RFI) environments, and search methods add additional challenges. Broad-band observations using a single instrument are not subject to these issues and can produce better results to distinguish between intrinsic causes and propagation effects \citep{Majid:2020}. The Ultra-Wideband Low (UWL) receiver system 
recently installed on the Murriyang, also known as the 64-m Parkes radio telescope \citep{Hobbs:2020} provides continuous frequency coverage in the band 704--4032~MHz with improved sensitivity over previous systems at the telescope. We have been using the UWL system to undertake a sensitive monitoring campaign of FRBs detected with the Australian Square Kilometre Array Pathfinder (ASKAP) telescope. The UWL is an excellent follow-up instrument with over ten times the instantaneous bandwidth of ASKAP and a factor of $\sim$~15  more sensitive.

FRB\,20190711A (hereafter referred to as FRB\,190711) was detected on 2019 July 11 at 01:53:41.09338~UTC with the ASKAP incoherent capture system (ICS) with a reported dispersion measure (DM) of $593.1\pm0.4$\DMunits and fluence ${34}\pm3$ Jy~ms \citep{190711ATel, Macquart:2020}. The incoherent detection triggered a voltage download that enabled interferometric localization of this FRB to a massive ($\sim 10^{9}$~M$_\odot$) star-forming galaxy \citep{Heintz:2020} at a redshift $z = 0.522$ \citep{Macquart:2020}. A study at high time and frequency resolution of the ASKAP data revealed several sub-bursts within its burst envelope with a complex dynamic spectrum which yielded a structure-optimized DM of 587.87\DMunits \citep{Day:2020}. Based on the CHIME population, \citet{Fonseca:2020} showed that repeating FRBs generally emit longer duration pulses relative to the one-off FRB sources. With a burst envelope width of $\sim 11 $~ms, the characteristic frequency drift in the dynamic spectrum and a flat polarization position angle (as a function of pulse phase), FRB\,190711 bears a strong resemblance to other repeating FRBs \citep{Day:2020}. 

In this paper, we report the discovery of a very narrow-band repetition from the source of FRB\,190711 using the UWL instrument. In Section~\ref{sec:observations}, we describe the observing campaigns and search strategies used for this FRB. In Section~\ref{sec:repeats}, we present the properties of the newly discovered repeat pulse. Finally, in Section~\ref{sec:discussion}, we discuss the implications for the FRB mechanism.

\section{Observations and Data Processing}\label{sec:observations}
\begin{figure}
    \includegraphics[width=\columnwidth]{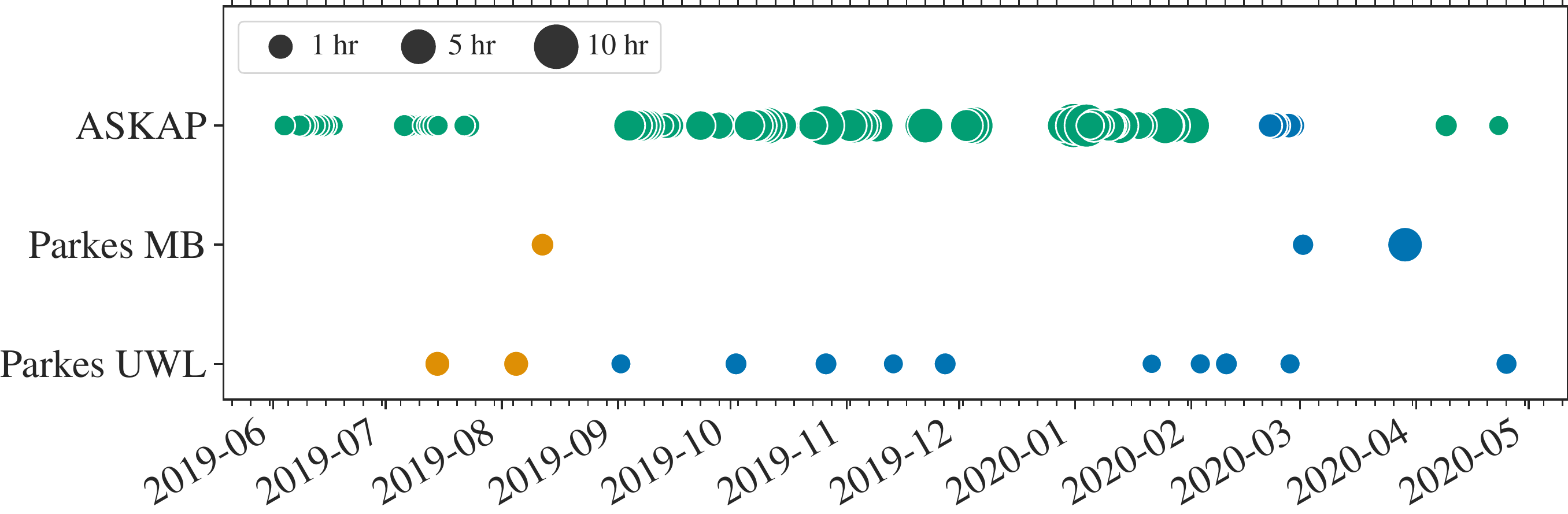}
\caption{Timeline of the follow-up observations for FRB\,190711. Observations of multi-beam position are shown in orange and of arcsecond-localized position in blue. ASKAP survey observations are shown in green.}
\label{fig:timeline}
\end{figure}

We used the 64-m Parkes radio telescope to follow up the position of the FRB\,190711 source. While our monitoring program of ASKAP-detected FRBs \citep{Kumar:2019, James:2020} also includes follow-up with the more sensitive GBT telescope, due to the southern circumpolar position of the source in the sky, we could not use it for FRB\,190711. Alongside Parkes, we also regularly observed the position of the FRB source with ASKAP\footnote{These observations include survey observations centred on the position RA=$22^{\rm h}$, Dec=$-80\degr$, and targeted observations of the burst source.}. The details of the follow-up observations and instruments used are listed in Table~\ref{tab:followupobs}. The majority of observations were centred at the interferometrically obtained arcsecond-localized position of the FRB source \citep{Macquart:2020}, i.e. RA = $21^{\rm{h}}57^{\rm{m}}40^{\rm{s}}$ and Dec = $-80\degr21\arcmin28\arcsec$ (J2000.0 epoch). Some of our initial observations were conducted at the less precise position reported in \citet{190711ATel}, which was based on a multiple-beam localization algorithm \cite[][]{Bannister:2017}. These multi-beam positions had an offset of ($\Delta_{\mathrm{RA}} = -4$ arcmin, $\Delta_{\mathrm{Dec}} = -2$ arcmin) from the interferometric position which is well within the primary beam (7 arcmin at 4.0~GHz) of Parkes. Figure~\ref{fig:timeline} shows a timeline of the follow-up observations of FRB\,190711. 

\begin{table}
\caption{FRB\,190711 follow-up observations.} 
\label{tab:followupobs}
\small
\centering
\begin{threeparttable}
\begin{tabulary}{1.08\columnwidth}{LCCCR}
\hline \hline
Instrument & Centre frequency & Bandwidth & Sensitivity\tnote{$\dagger$} & Obs.\\
& (MHz) & (MHz) & (Jy ms) & (h)\\
\hline 
ASKAP ICS  & 864--1320 & 336 & 3.7 $N_{\mathrm{ant, 36}}^{-0.5}$ & 292.9 \\
Parkes MB & 1382 & 340  & 0.5 & 8.1  \\
Parkes UWL & 2368 & 3328  & 0.15 $\Delta \nu_{\mathrm{width, 3.3}}^{-0.5}$ & 11.0\\
\hline
\end{tabulary}
\begin{tablenotes}[flushleft]
    \item[$\dagger$] The limiting fluence for a pulse width of 1~ms and S/N threshold of 10$\sigma$. $\Delta \nu_{\mathrm{width, 3.3}}$ is the burst emission width in units of 3.3~GHz.
\end{tablenotes}
\end{threeparttable}
\end{table}  

\subsection{ASKAP searches}
Both the observations of the FRB\,190711 field and the targeted follow up of the FRB source were conducted using ASKAP incoherent sum mode, in which all antennas are pointing to the same location and the intensities from each antenna are added (making the sum incoherent). The data are searched in near-real-time (latency $\lesssim 1$~s) using the custom GPU-based detection pipeline {\sc fredda} \citep{fredda_ascl}. A description of the detection methods and search pipeline can be found in the supplementary materials in \citet{Bannister:2019_localization}. We found no other astrophysical events exceeding a threshold signal-to-noise ratio (S/N) of 10 in 293 h of observation.

\subsection{UWL searches}
We used the UWL receiver at Parkes, covering a continuous frequency range from 704 to 4032~MHz. 
Input signals for the system are digitized and recorded to produce 26 contiguous sub-bands, each with a bandwidth 128~MHz. In our observations, the sub-band data were sampled with a time resolution of 64~$\upmu$s in 256 frequency channels with each channel coherently dedispersed using a convolving algorithm \citep{Hobbs:2020}. The data were then combined at a later stage and stored in an 8-bit sampled {\sc psrfits} search-mode file \citep{psrchive} with four polarization products. 

A standard search pipeline forms a time-series by summing the whole frequency band at several DM trials and then searches for pulses in it. However, in our case, given the wide bandwidth of the UWL system and the observed low spectral occupancy in the emission from many FRB sources \citep{Shannon:2018, CHIME_repeaters19_2, Gourdji:2019}, summing the whole 3.3~GHz band was likely to be suboptimal. We therefore, conducted a comprehensive search of this wide-band data by dividing into sub-bands of different sizes and searching each sub-band independently. We successively searched the data, sub-banding into widths of size 1$\times$3328 MHz, 2$\times$1664 MHz, 4$\times$832 MHz, 13$\times$256 MHz, 26$\times$128 MHz, and 52$\times$64 MHz. For each sub-band width, we also searched overlapping adjacent sub-bands by shifting the bands by half the sub-band width to capture signal overlapping a boundary.

\begin{figure}
	\centering
\includegraphics[width=0.98\columnwidth]{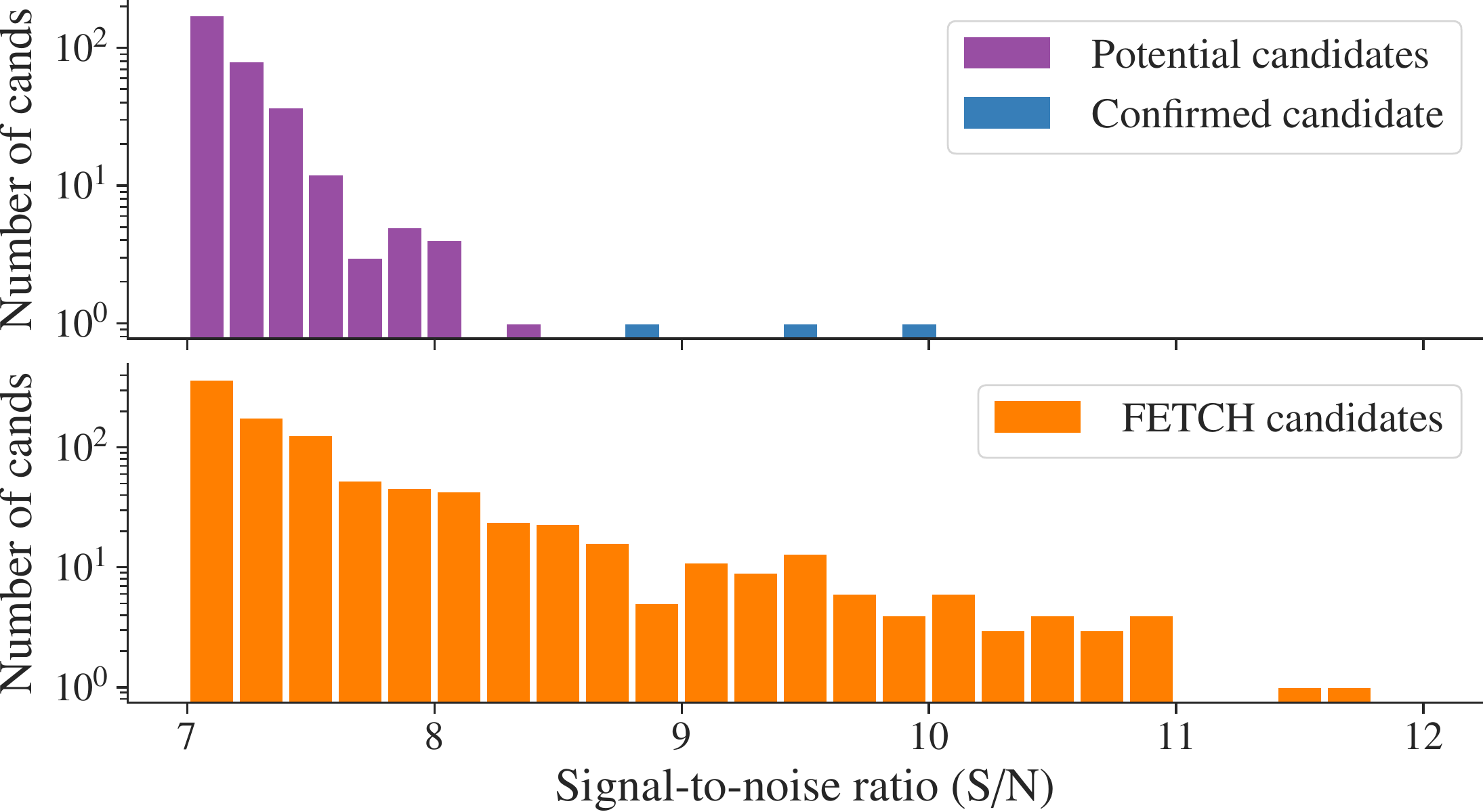} 
\caption{S/N distribution of single-pulse candidates. {\em Bottom panel}: {\sc fetch} classified candidates with probability > 0.5. {\em Top panel}: {\sc fetch} candidates after further manual vetting.  We note that, due to the iterative search method, unique candidates can appear more than once. In this case, the repeat burst has been detected in three overlapping sub-band searches.}
\label{fig:fetch_stats}
\end{figure}

We first formed a total intensity {\sc sigproc}\footnote{\url{http://sigproc.sourceforge.net}} filterbank format file from the {\sc psrfits} data. Using statistical moments (variance, skewness and kurtosis) of each channel and their median absolute deviation statistics, we used a modified $Z$-score threshold of 5 to identify and flag channels affected by RFI. Then, we searched each sub-band independently for dispersed pulses using the GPU based single-pulse search software {\sc heimdall\footnote{\url{https://sourceforge.net/projects/heimdall-astro/}}} \citep{Barsdell:2012PhDT}. The pipeline searched over a DM range of 100 to 1000\DMunits using a tolerance of 25, 10, 5, 0.5, 0.2 and 0.1 per cent for each iteration of the search (in the decreasing order of sub-band width from 3328 to 64 MHz), respectively. We then applied the following criteria to filter the clustered candidate list obtained from {\sc heimdall}: S/N $\geq 7.0 $, $0.128$~ms $ \leq $ pulse width $\leq 32.768$~ms and members\footnote{Number of individual boxcar/DM trials clustered into a candidate.} $> 10 $. This resulted in a total of 59407 candidates for 11 h of observation. We note that the filtering criteria used for pulse width are consistent with the observed values for FRBs \citep{Fonseca:2020}.

We used the convolutional neural network {\sc fetch\footnote{\url{https://github.com/devanshkv/fetch}}} \citep{Agarwal:2020} to perform the FRB/RFI binary classification of candidates. We used model {\tt A} with a probability threshold of 0.5 and obtained 948 potential candidates. Due to a large number of false positives, we do not include other {\sc fetch} models in this analysis and use only the best model. We then visually inspected each of the classified candidates and found one astrophysical burst at a DM~$\sim$~587\DMunits. We found the burst in three of the overlapping sub-band searches. In Figure~\ref{fig:fetch_stats}, we show the S/N distribution of all candidates labelled positive by {\sc fetch}. To test the reliability of {\sc fetch}, we also visually inspected all the 3100 candidates found in the DM range of 545 to 625\DMunits, as any repeat bursts from the source would have a similar DM to the earlier FRB\,190711 burst, and found no other astrophysical pulses.

\subsection{Parkes multibeam searches}
In some epochs (see Figure~\ref{fig:timeline}), typically when the UWL was not available, we used the 20-cm multibeam (MB) receiver at Parkes to search for bursts from the FRB\,190711 source. An overview and details of the data format, detection methods and search pipeline may be found in \citet{Kumar:2019} and references therein. We found no bursts of astrophysical origin in 8 h of observation.

\begin{table}
\caption{Measured properties of the repeat burst from the FRB\,190711 source.}
\label{tab:burstsproperties}
\centering
\begin{threeparttable}
\begin{tabulary}{0.99\columnwidth}{LC}
\hline \hline
Parameter & Value\\
\hline
Event identifier  & FRB 20190804A \\
Arrival time (UTC)\tnote{(1)} & 2019-08-04 19:54:29.9263(1) \\
Arrival time (MJD)\tnote{(1)} & 58699.829513035(1) \\
Dispersion measure (\DMunits)\tnote{(2)} & $587.4^{+1.7}_{-2.7}$\\
Pulse width (ms) & $1.0 \pm 0.1$ \\
Spectral width (MHz) & $65 \pm 7$ \\
Centre frequency of emission (MHz) & $1355 \pm 3$ \\
Integrated S/N  & 11.7 \\
Peak flux density (Jy)\tnote{(2)} & $1.4 \pm 0.2$ \\
Fluence (Jy ms) & $1.4 \pm 0.1$ \\
Spectral energy density (erg~Hz$^{-1}$) & $6.8 \times 10^{30}$\\
\hline
\end{tabulary}
\begin{tablenotes}[flushleft]
    \item[(1)] {Burst time of arrival is referenced to 1375 MHz, and the uncertainties are in parentheses.} 
    \item[(2)] {DM error ranges correspond to an uncertainty of one in S/N. The uncertainties on flux density correspond to the rms noise for the burst.}
\end{tablenotes}
\end{threeparttable}
\end{table}

\section{The Repetition}\label{sec:repeats}
\begin{figure*}
\begin{center}
  \includegraphics[scale=0.40]{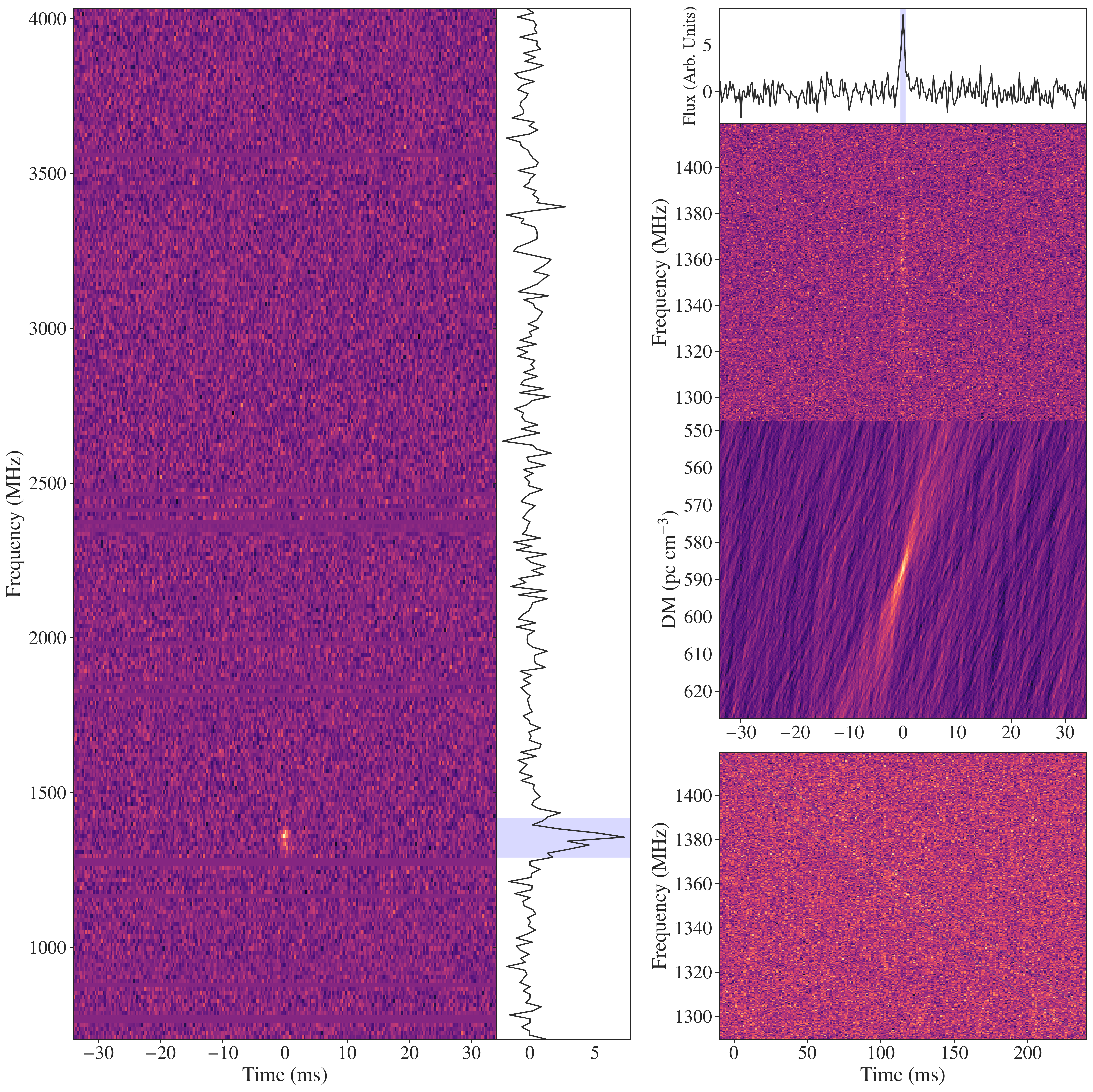}
\end{center}
\caption{
Dynamic spectra and diagnostic plots of the UWL detection from FRB\,190711 source. Data have been dedispersed to the best-fitting DM of 587.4\DMunits. {\em Left subplot}: The left-hand panel shows the dynamic spectrum (frequency resolution = 26~MHz, time resolution = 0.26~ms), and the right-hand panel shows the time-averaged on-pulse spectrum across the full UWL band. {\em Right subplot}:
The top panel shows the dynamic spectrum over the frequency range of the best-fitting sub-band in the second row (frequency resolution = 0.5~MHz) alongside the frequency-averaged flux density in the first row and the DM-time image in the third row. The bottom panel is the dynamic spectrum prior to dedispersion.
\label{fig:repeaterplots}  }
\end{figure*}

The repeat burst from the source of FRB\,190711 that we identified in the UWL observations occurred 24 days after the initial ASKAP detection. The dynamic spectra of the repeat burst, along with the DM-time transform and the on-pulse spectrum are shown in Figure~\ref{fig:repeaterplots}.

The most striking feature of the burst is the absence of signal in nearly all of the observed bandwidth, demonstrating the futility of integrating over the entire band to search for and characterize this burst. We identified the portion of the spectrum where the burst is bright, by fitting the on-pulse average spectrum with a Gaussian function and extracting a band twice the measured full-width at half-maximum (FWHM) around the peak signal (shown as a shaded region in Figure~\ref{fig:repeaterplots}). Unless otherwise mentioned, the results reported here are based on the obtained sub-band of bandwidth 130 MHz. We obtain a best-fitting DM of 587.4\DMunits\,by maximizing the S/N, which is consistent with the DM estimated for the original ASKAP detection \citep{Day:2020}. We fitted the frequency-averaged pulse profile with a single Gaussian function to measure the temporal width of the burst and estimate an FWHM width = 1~ms. Thus, we find an order of magnitude difference in the widths of the pulses detected to date for FRB\,190711. We do not find any temporal sub-structure in the dynamic spectrum of the repeat burst. We do not attempt to measure a scattering time-scale given the low S/N of the burst. We estimate the integrated S/N to be $\sim$~12 by averaging over the time bins within twice the FWHM pulse width. The burst properties are listed in Table~\ref{tab:burstsproperties}. 

\subsection{Burst polarimetry}
We formed a full-Stokes parameter {\sc psrfits} format archive file for the repeat burst by extracting the UWL data using {\sc dspsr} \citep{dspsr}. We then calibrated the archive-format data to measure the flux and polarimetric properties using procedures detailed in \citet{Lower:2020} and references therein. For polarization calibration, we used a short observation (2.3 min) of a linearly polarized noise diode, which was obtained at the start of the observing session during which the repeat burst was detected.

\begin{figure}
	\centering
\includegraphics[width=0.99\columnwidth]{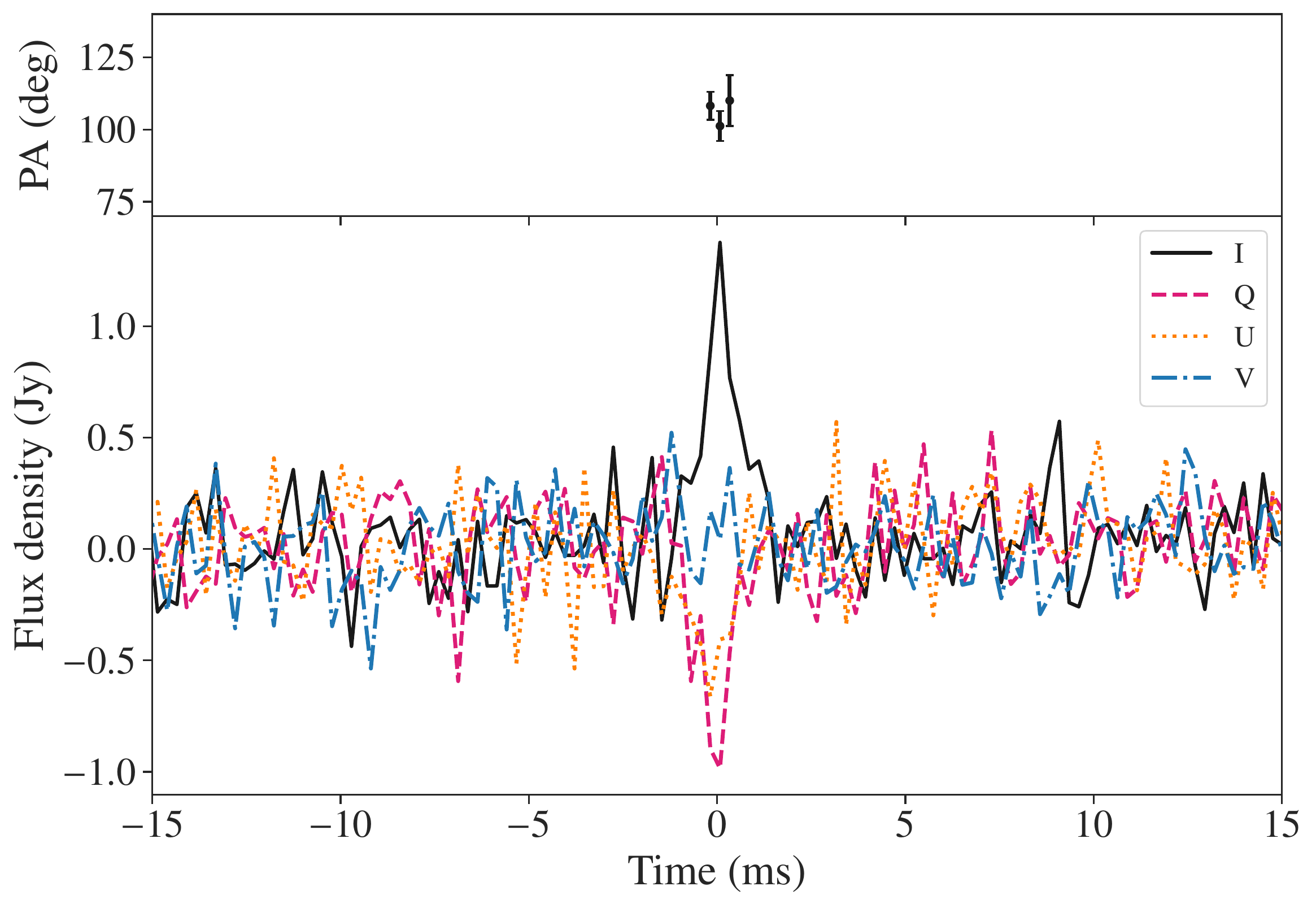} 
\caption{Polarization profile of the repeat burst from the FRB\,190711 source. {\em Top panel}: Position angle (PA) versus time. {\em Bottom panel}: Frequency-averaged time series for the four Stokes parameters. The data are corrected for the ASKAP-measured RM of 9\RMunits.}
\label{fig:polprofile}
\end{figure}

We attempted to search for Faraday rotation in the Stokes data using the {\sc rmfit} routine from {\sc psrchive} but could not constrain the rotation measure (RM) given the relatively low signal-to-noise ratio of the burst. We then corrected for Faraday rotation using the RM reported by \citet{Day:2020} for the initial ASKAP-detected burst and saw no significant change in the burst polarization. We also determined the absolute polarization position angle (PA) using the frequency-averaged Stokes Q and U. We did not de-bias the total linear polarization. Instead, we used a $ 3\sigma $ threshold on Stokes I to mask noise values. Figure~\ref{fig:polprofile} shows the frequency-averaged Stokes profile of the repeat burst along with the PA.  

The burst has a fractional linear polarization of $\sim 0.8$ and a flat PA with mean $107 \pm 4 \degr $ as a function of pulse phase. The S/N of the pulse is not sufficiently significant to infer further information. For the ASKAP-detected burst, \citet{Day:2020} found no evidence of any circular polarization and a pulse-averaged linear polarization of $\sim 100$ per cent. The polarization angle was also found to be flat. Using the ASKAP data, we obtain a mean value of $83 \pm 1 \degr$ for the position angle.

\subsection{Spectral properties} 
We formed the time-averaged pulse spectrum over the extent of twice the measured FWHM of the pulse width. The spectrum is highly band-limited with the emission peaking around 1355 MHz and showing a fractional FWHM emission bandwidth of  $\approx$ 2 per cent of the observing UWL band. The emission band ($\sim $ 130 MHz) of the repeat burst overlaps roughly with the top half of the band observed for the original ASKAP detection. The overlapping band starts just above the high-end cut-off $\sim$ 1300~MHz of the ASKAP spectrum.

\begin{figure}
	\centering
\includegraphics[width=0.99\columnwidth]{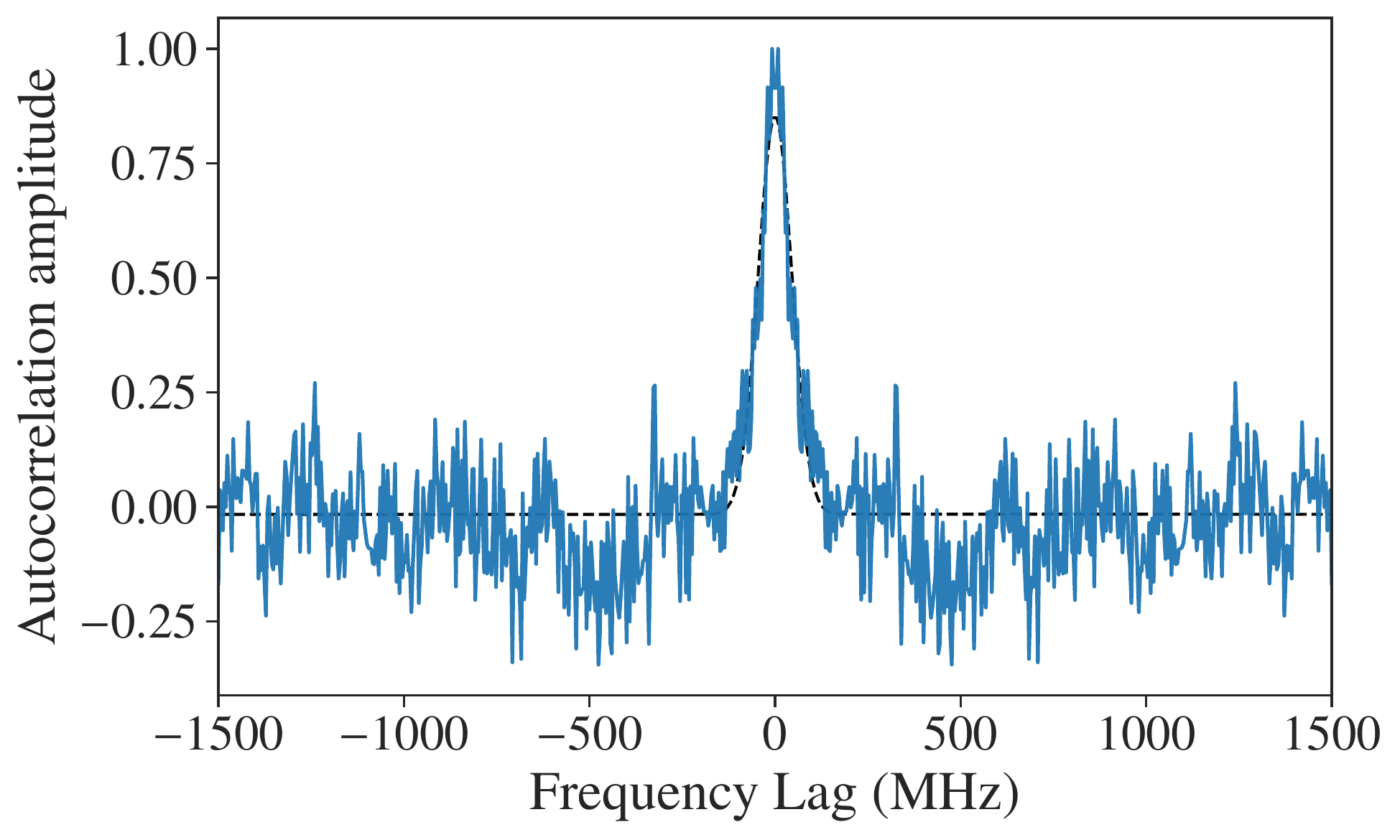} 
\caption{Autocorrelation function of the time-averaged spectrum of the repeat burst (resolution $= 4$~MHz). The zero lag noise spike has been removed. The best-fitting model using a single-component Gaussian is shown as a dashed black line.}
\label{fig:acfspectra_uwl}
\end{figure}

No pulse broadening is observed in the original ASKAP detection \citep{Qiu:2020}. We use an autocorrelation function (ACF) analysis \citep{Farah:2018} to measure the scintillation bandwidth in the repeat burst and compare it with that of the ASKAP detection. The scintillation bandwidth is defined as the half-width-half-maximum (HWHM) of a Gaussian fitted to the ACF \citep{Spitler:2018}. The ACF of the repeat burst (see Figure~\ref{fig:acfspectra_uwl}) is best described by a single characteristic frequency scale of HWHM 53~MHz. To measure scintillation bandwidth in the ASKAP spectrum, we use the low frequency-resolution data with 1~MHz channel bandwidth and find two characteristic frequency scales of HWHM 4.6 and 90~MHz.

We measure the integrated S/N for the rest of the UWL band by forming sub-bands of size 130~MHz. The pulse window is fixed for each sub-band based on the narrow-band detection using time bins within twice the FWHM width. No signal above $6 \sigma$ is detected in any of the other sub-bands. We place an upper limit of 0.4 $(\Delta \nu_{\mathrm{width}}/130\, \mathrm{MHz})^{-0.5}$~Jy\,ms on the fluence of the burst emission at 0.7--1.3 and 1.4--4.0~GHz (with an emission width of $\Delta \nu_{\mathrm{width}}$) during this observation, assuming a nominal pulse width of 1~ms.

\section{Discussion and  Conclusions}\label{sec:discussion}
At a redshift of $z = 0.522$, the FRB\,190711 source is the most distant and produces the most luminous bursts of the repeating FRBs identified to date. The rotation measure of this source is 9\RMunits \citep{Day:2020} much smaller in magnitude compared to other repeating sources \citep{Michilli:2018, Fonseca:2020}. The repeat pulse has a flat PA and is linearly polarized, similar to the original burst. The ASKAP detection showed a sub-pulse drift of $\sim 15$~MHz\, ms$^{-1}$. We do not see any frequency drift in the repeat burst despite the adequate frequency and time resolution, which might be explained by the low S/N of the burst. The UWL detection is $\sim$ 25 times fainter than the original ASKAP detection. The lack of emission in > 3\,GHz of the UWL band provides further evidence for having preferred frequencies of emission \citep{Gourdji:2019}, as has also been suggested for the periodic repeater source FRB\,20180916B \citep{Aggarwal:2020RNAAS}.

\subsection{Explaining the burst spectral structure} 
The NE2001 line of sight model \citep{cordes2001model} predicts a scattering time-scale of 0.14~$\upmu$s and a scintillation bandwidth of $\approx$~1.3~MHz at 1.4~GHz from diffractive scintillation from the Milky Way. We note that due to low S/N and \mbox{0.5-MHz} spectral resolution, the UWL data are insufficient to resolve the predicted width. 

The spectral structure is inconsistent with diffractive interstellar scintillation (DISS). Even if the emission represents the brightest scintle within the $3.3$~GHz bandwidth, we would expect underlying emission to be present throughout the band that would be detected when integrated over the band. However, a comparison of the bandwidths of the ASKAP and Parkes bursts is consistent with propagation through turbulent plasma, in which spectral scales are proportional to $\Delta \nu_{\mathrm{DISS}} \propto \nu^{-4}$. If we attribute the spectral structure scale (centred at 1190 MHz) in the ASKAP spectrum to DISS, then we would expect the band extent of the emission (centred at 1355 MHz) in the UWL repeat burst to be $(\nu_{\rm UWL}/\nu_{\rm ASKAP})^{-4} \times 90~{\rm MHz} \approx 54 $ MHz. This is consistent with the characteristic emission scale of 53~MHz observed in the repeat burst.
We note that the lower end of the ASKAP spectrum cannot be precisely determined due to the limited observed bandwidth, so we have assumed that it cuts off close to the bottom of the ASKAP band. Further bursts from this source will better constrain a propagation model.

It is possible that the spectral structure could be the result of plasma lensing. \citet{Cordes:2017} show that caustics due to plasma lensing can produce strong magnifications ($\lesssim 10^{2}$) on short time-scales. These magnifications appear as narrow spectral peaks (0.1--1.0~GHz) in the burst spectra even if the intrinsic spectrum is smooth and broadband. These multiple burst images can also interfere and produce frequency structures on scales of $ \sim $~1--100~MHz. Depending on the properties and geometrical complexity of the lens, there may be multiple-peaked gains in the observed spectrum. We neither see multiple peaks nor the double-peaked gain cusps \citep{Law:2017} predicted for a single Gaussian lens over the observed bandwidth of $ 3.3$~GHz. Based on the detection of a single spectral component, we are not able to determine the existence of a focal frequency in our observed bandwidth. Further wide-band spectra of this source ($>\,$4~GHz) would be necessary to determine the focal frequency and lens parameter, thus providing constraints on the effects of plasma lensing for bursts from this FRB source. We note that a key prediction from this model is that there is no directional dependence of the sub-pulse drifts. However, thus far, only downward drift \citep{Hessels:2019} is seen in repeating FRBs, including the original burst from the FRB\,190711 source. 

Another possibility is that a mechanism intrinsic to the FRB source can explain both the limited emission bandwidth of FRBs and downward drifting in sub-pulses. One such proposed mechanism is the synchrotron maser emission model from decelerating blast waves \citep{Metzger:2019}. This model also naturally accounts for the high linear polarization fraction and the high efficiency to produce coherent radiation \citep{Margalit:2020}. The model suggests emission to be narrowly peaked in frequency due to the combined effects of induced scattering at lower frequencies and the fall-off of the intrinsic maser emission at high frequencies. However, the predicted emission width $ \Delta \nu/\nu \sim 1$ is more than an order of magnitude larger than what we observed $ \Delta \nu/\nu \sim 0.05$ in the UWL detection. 

A final possibility is that there is further emission below our detection threshold. We can have an intrinsically frequency-dependent spectrum from the burst source such that only the signal in the brightest parts is above the threshold noise. Thus the rest (either broadband or multiple brightness peaks) portion of the intrinsic spectrum is not visible. In this case, the occupancy of other bright patches would have to be sufficiently low to not be detected as a broad-band signal.  

\subsection{Finding narrowband bursts}
The detection of an extremely band-limited burst suggests a re-think of the conventional FRB search methods and motivates the implementation of multi-bandwidth burst search strategies. The burst would not have been detected if we had not searched in frequency bandwidths $\sim 100$~MHz, suggesting such sub-band methods will play an increasingly large role in FRB searches with wide-band instruments. As such searches will significantly increase the number of potential candidates, further development of machine-learning-based classifiers and optimization of sub-band search strategies will be vital in finding more band-limited transients from FRB sources. With upcoming more broad-band systems like UWL will open up the opportunity to increase the detection rate of FRBs if emission is as narrow as seen from the source of FRB\,190711. 

\section*{Acknowledgements}
PK acknowledge support through the Australian Research Council (ARC) grant FL150100148. 
RMS acknowledges support through ARC grants  DP180100857 and FT190100155. 
ATD is the recipient of an ARC Future Fellowship (FT150100415).
This work was performed on the OzSTAR national facility at Swinburne University of Technology. The OzSTAR program receives funding in part from the Astronomy National Collaborative Research Infrastructure Strategy (NCRIS) allocation provided by the Australian Government. Work at NRL is supported by NASA. The Parkes Radio Telescope and the Australian Square Kilometre Array Pathfinder are part of the Australia Telescope National Facility which is managed by CSIRO. Operation of ASKAP is funded by the Australian Government with support from the NCRIS. ASKAP uses the resources of the Pawsey Supercomputing Centre. Establishment of ASKAP, the Murchison Radio-astronomy Observatory (MRO) and the Pawsey Supercomputing Centre are initiatives of the Australian Government, with support from the Government of Western Australia and the Science and Industry Endowment Fund. We acknowledge the Wajarri Yamatji people as the traditional owners of the MRO site and the Wiradjuri people as the traditional owners of the Parkes observatory site.

\vspace{-0.4cm}
\section*{Data Availability}

The data underlying this article will be shared on reasonable request to the corresponding author.  Raw data from the Parkes telescope are archived on the CSIRO data access portal \url{https://data.csiro.au}. 



\bibliographystyle{mnras}
\bibliography{references} 






\bsp	
\label{lastpage}
\end{document}